\documentclass[twocolumn,amsmath,amssymb,mathabx,floatfix,pra]{revtex4}

\usepackage{graphicx}			% Include figure files

\def\nm{{\ {\rm nm}}}						% nm
\def\micron{{\ \mu{\rm m}}}					% microns
\def\Hz{{\ {\rm Hz}}}						% Hz
\def\kHz{{\ {\rm kHz}}}						% kHz
\def\Er{{{E_r}}}							% Er
\def\kr{{{k_r}}}							% kr
\def\Rb87{^{87}\rm{Rb}}						% Rb 87
\def\Li6{^{6}\rm{Li}}						% Li 6
\newcommand{\ket}[1]{|#1\rangle}

\begin{document}

\title{Raman processes and effective gauge potentials}

\author{I.~B.~Spielman}
\affiliation{Joint Quantum Institute, National Institute of Standards and Technology, and University of Maryland, Gaithersburg, Maryland, 20899, USA}

\date{\today}

\begin{abstract}
A new technique is described by which light-induced gauge potentials allow systems of ultra-cold neutral atoms to behave like charged particles in a magnetic field.  Here, atoms move in a uniform laser field with a spatially varying Zeeman shift and experience an effective magnetic field.  This technique is applicable for atoms with two or more internal ground states.  Finally, an explicit model of the system using a single-mode 2D Gross-Pitaevskii equation yields the expected vortex lattice.
\end{abstract}

\maketitle

\section{Introduction}

Condensed matter systems are replete with many-body effects, where the interactions between the innumerable particles determine the basic physics of the system.  Of late, ultracold atoms have demonstrated a range of basal condensed matter systems and effects: Bose-Einstein condensation (BEC) ~\cite{Anderson1995,Davis1995a}, Tonks-Girardeau gases~\cite{Kinoshita2004,Paredes2004}, the superfluid to Mott-insulator transition~\cite{Greiner2002}, Berezinskii-Kosterlitz-Thouless physics~\cite{Hadzibabic2006} in bosons; and the crossover from a Bose condensate to a Bardeen-Cooper-Schrieffer paired superfluid in fermions~\cite{Greiner2005a}.  Here, I discuss a technique to realize more complicated states where the charge neutral atoms behave as charged particles in a magnetic field.

In a magnetic field a 2D electron gas (2DEG), can display a range of exotic phenomena, including the integer quantum Hall effect (IQHE) and fractional quantum Hall effect (FQHE).  The FQHE states are exotic quantum liquids, where the lowest energy charged excitations are fractionally charged quasiparticles.  While remarkable, ideas of charge and spin fractionalization are now well established in modern descriptions of strongly interacting quantum systems.  More recently, exotic -- non-Abelian -- states useful for topological quantum computation have been predicted, but remain experimentally elusive~\cite{Nayak2008}.  Still, experimental evidence now strongly supports the existence of fractionally charged excitations in these systems~\cite{de-Picciotto1997}, but evidence for their statistics is less conclusive~\cite{Camino2007,Radu2008}. 

Systems of ultracold atoms are uniquely positioned to realize topological phases arising in strongly interacting 2D systems, Fermi and Bose alike.  The more exciting FQHE states are inevitably very delicate and only exist in the most clean systems, if at all.  As with a 2DEG, the states of the system can be labeled by the filling factor $\nu = n/\Phi$, the ratio of the 2D particle density $n$ to magnetic flux $\Phi=eB/h$.  When $\nu\lesssim1$ a 2DEG can display a range of exotic phenomena, including various FQHE states.  The primary challenge is to engineer a Hamiltonian for which neutral atoms behave as charged particles in a magnetic field. 

\begin{figure}[tbp]
\begin{center}
\includegraphics[width=3.3in]{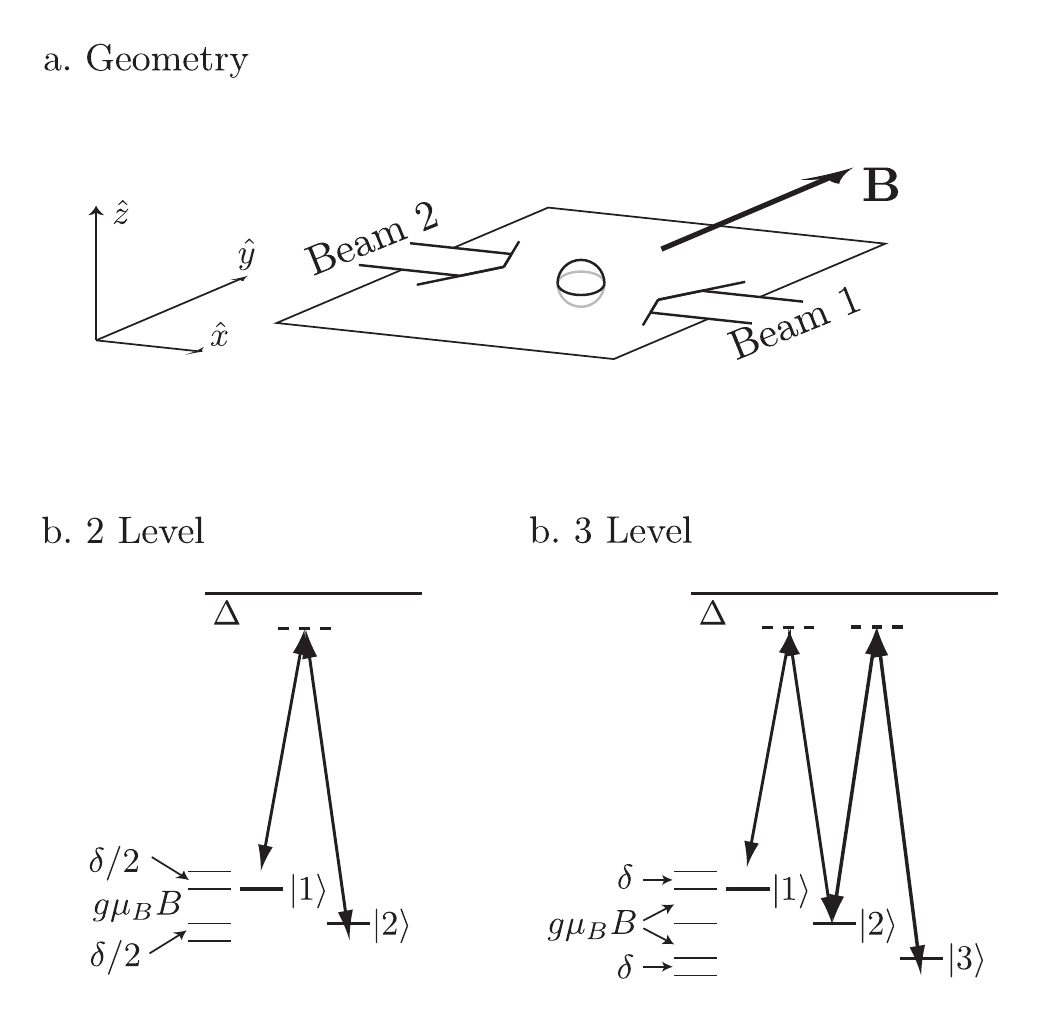}
\end{center}
\caption[Geometry of dressed state laser system]{
Geometry and level diagram to realize light-induced gauge potentials.  a. The figure depicts the two counter-propagating laser beams with momentum ${\bf k_1}$ and ${\bf k_2}$, aligned along $+\hat x$ and $-\hat x$ respectively, and the physical magnetic field along $\hat y$.  b. Level diagram for two electronic ground states. c. Level diagram for three electronic ground states; a small quadradic Zeeman shift $\epsilon$ of state $\ket{2}$ is not depicted.  Additional transitions, relevant for physical atoms, contribute to the state-independent light shift in this configuration.
}
\label{geometry}
\end{figure}

This paper describes a new procedure for creating light-induced gauge potentials~\cite{Juzeliunas2006,Zhu2006,Liu2007,Cheneau2008,Gunter2009}, as a way to create an effective magnetic field.  This work focuses on the adiabatic eigenstates of atoms in the presence of two optical fields, and finds that the resulting Hamiltonian can describe charged particles in a magnetic field.  In contrast to earlier proposals, the requisite spatial inhomogeneity here is provided by an external magnetic field gradient instead of inhomogenous optical fields.  This paper first presents explicit results for a model system with two coupled states, and then generalizes to the three-level case (relevant to the $F=1$ manifold of $\Rb87$).  The effective magnetic field resulting from light-induced gauge potentials exists in a small spatial region.  An important finding is that coupling between more than two internal states increases the spatial range over which large effective fields can be realized.
  
In second-quantized form, the Hamiltonian for a particle in a uniform time-independent magnetic field normal to a 2D plane is
\begin{align}\label{FieldHam}
\int d^2 {\bf x}\frac{\hbar^2}{2 m}\hat \psi^\dagger({\bf x}) \left\{\left[k_y - \frac{q A_y}{\hbar} \right]^2 + \left[k_x - \frac{q A_x}{\hbar} \right]^2\right\}\hat \psi({\bf x}),
\end{align}
where $\psi^\dagger({\bf x})$ is the field operator for the creation of a particle at ${\bf x}$, and $k_{x,y}=-i\partial_{x,y}$.  For real magnetic fields the vector potential has gauge freedom.  For example the Landau gauge choice, $\left\{A_x = B y, A_y = 0\right\}$, gives a uniform magnetic field along $\hat z$.  This proposal explicitly realizes Eq.~\ref{FieldHam} in a specific gauge (the Landau gauge for the geometry discussed below): a Hamiltonian where the minimum of the energy-momentum dispersion relation $E({\bf k})$ becomes asymmetric~\cite{Higbie2002} and is displaced from zero momentum as a function of spatial position.  The dressed single-particle states are spin/momentum superpositions whose state decomposition depends on the local value of the effective vector potential ${\bf A}$.  In this way, the canonical momentum associated with the Landau gauge is physically observable by probing the internal-state decomposition of the adiabatic dressed states.  This effective Landau-gauge vector potential was recently measured by Lin et al~\cite{Lin2009a}.

Our approach relies on a collection of bosons with two or more relevant electronic ground states interacting with two counter-propagating ``Raman'' lasers aligned along $\hat x$ that are detuned from each other by $\Delta_{R}$, shown in Fig. \ref{geometry}.  A small magnetic field ${\bf B} = (B_0 + \Delta B)\hat y$, introduces a linear Zeeman splitting between the levels;  $\Delta_R = g \mu_{\rm B} B_0$, so $\delta = g \mu_{\rm B}\Delta B$ is the detuning from Raman rasonance. Here $g$ is the atomic $g$-factor, and $\mu_B$ is the Bohr magneton.  I focus on the limit when both Raman beams are far detuned from the ground to excited state transition so there is negligible population in the excited state, and the Raman beams simply induce a coupling $\Omega$ between ground states.  As shown below, this set of coupling fields can lead to effective magnetic fields.

\section{Two-Level System}

For simplicity, first consider a two level system with internal states $\ket{+}$ and $\ket{-}$, where exact solutions, studied in the context of a 3D BEC in Ref.~\cite{Higbie2002}, are readily available.  (Physically, these two states might be two $m_F$ levels in the ground state manifold of an alkali atom; for example, the $F=1$ manifold of $^{87}{\rm Rb}$ at large enough field that the quadratic Zeeman effect resolves the three Zeeman sublevels.)  Since counter-propagating Raman beams aligned along $\hat x$ couple states differing in $k_x$ by $2\kr$, the recoil momentum $\kr=2\pi/\lambda$ and energy $\Er = \hbar^2 \kr^2/ 2 m$ will be taken as the units of momentum and energy.  Here, $\lambda$ is the wavelength of the nearly degenerate Raman beams, $m$ is the atomic mass, and the two-photon Raman coupling is $\Omega$.  In the frame rotating at $\Delta_R/h$ the Raman fields are detuned $\delta= g\mu_B \Delta B$ from resonance and the atom-light coupling term in the rotating wave approximation (RWA) is
\begin{align*}
\hat H' =&\int d y \int\frac{dk_x}{2 \pi}\left\{\frac{\Omega}{2}\left[\hat \phi_+^\dagger(k_x-2,y)\hat \phi_-(k_x,y) + {\rm H.C.}\right]\right. \\
& +\left.\frac{\delta}{2}\left[\hat\phi_+^\dagger(k_x,y)\hat \phi_+(k_x,y) - \hat\phi_-^\dagger(k_x,y)\phi_-(k_x,y)\right]\right\}
\end{align*} 
The notation $\hat \phi_\sigma^\dagger(k_x,y)$ denotes the creation of a particle with wave vector $k_x$ along $\hat x$ at position $y$, with $\sigma=\pm$ and H.C. indicates the Hermitian conjugate.  Also, observe that $\hat H'$ includes the Raman detuning terms.  In the following analysis $\Omega$ and $\delta$ will be treated as spatially varying functions of $y$, but not $x$.

Absent coupling, the Hamiltonian for particles in 2D is a sum $\hat H =\hat H_x + \hat H_y + \hat V +\hat H_{\rm int}$.  Respectively, these represent motion along $\hat x$, motion along $\hat y$, the external potential, and interparticle interactions.  When expressed in terms of the real space field operators $\hat \psi_\sigma({\bf r})$, these terms are
\begin{align*}
\hat H_x &= \int d^2{\bf r}\sum_\sigma\hat\psi^\dagger_\sigma({\bf r})\left[-\partial_x^2\right] \hat\psi_\sigma({\bf r})\\
\hat H_y &= \int d^2{\bf r}\sum_\sigma\hat\psi^\dagger_\sigma({\bf r})\left[-\partial_y^2\right]\hat\psi_\sigma({\bf r})\\
\hat V &=  \int d^2{\bf r}\sum_\sigma\hat\psi^\dagger_\sigma({\bf r})V({\bf r})\hat\psi_\sigma({\bf r})\\
\hat H_{\rm int} &= \frac{g_{\rm 2D}}{2} \int d^2{\bf r}\sum_{\sigma,\sigma'} \hat\psi^\dagger_\sigma({\bf r}) \hat\psi^\dagger_{\sigma'}({\bf r}) \hat\psi_\sigma({\bf r}) \hat\psi_{\sigma'}({\bf r}).
\end{align*}
The contact interaction for collisions between ultra-cold atoms in 3D is set by the 3D s-wave scattering length $a_s$, here assumed to be state independent.  Strong confinement in one direction yields an effective 2D coupling constant $g_{\rm 2D}=\sqrt{8\pi}\hbar^2 a_s/m l_{\rm HO}$.  $l_{\rm HO}$ is the harmonic oscillator length resulting from a strongly confining potential along $\hat z$; a 1D optical lattice, for example~\cite{Petrov2001}.  Finally, $V({\bf r})$ is an external trapping potential, also taken to be state-independent.

This problem is exactly tractable when considering free motion along $\hat x$, i.e., treating only $\hat H_x$ and $\hat H'$.  The second quantized Hamiltonian for these two contributions can be compactly expressed in terms of the operators $\{\hat \varphi^\dagger_+(\tilde k_x,y),\hat \varphi^\dagger_-(\tilde k_x,y)\} = \{\hat \phi^\dagger_+(\tilde k_x-1,y),\hat \phi^\dagger_-(\tilde k_x+1,y)\}$.  Using this Nambu spinor, $H\approx\hat H_x+\hat H'$ reduces to a integral over 2$\times$2 blocks 
\begin{align}
H({\tilde k}_x,y) &=
\left(
\begin{array}{cc}
(\tilde k_x - 1)^2 + \delta/2 & \Omega/2 \\
\Omega/2 & (\tilde k_x + 1)^2 - \delta/2 
\end{array}\right)\label{TwoByTwo}
\end{align}
labeled by $\tilde k_x$ and $y$.  The dependence of the two-photon coupling $\Omega$ and detuning $\delta$ on $y$ has been suppressed for notational clarity.  The resulting Hamiltonian density for motion along $\hat x$ at a fixed $y$ is 
\begin{align*}
\mathcal{H}(y) & = \int\frac{d \tilde k_x}{2\pi} \sum_{\sigma,\sigma'} \hat \varphi^\dagger_\sigma(\tilde k_x,y) H_{\sigma, \sigma'}(\tilde k_x,y) \hat \varphi_{\sigma'}(\tilde k_x,y).
\end{align*}
For each $\tilde k_x$, $H(\tilde k_x,y)$ can be simply diagonalized into spin-momentum superposition states by the unitary transformation $U(\tilde k_x,y) H(\tilde k_x,y) U^\dagger(\tilde k_x,y)$.  The resulting eigenvalues $E_\pm(\tilde k_x,y)=\tilde k_x^2+1\pm\sqrt{(4 \tilde k_x - \delta)^2 + \Omega^2}/2$ give the effective dispersion relations in the dressed basis $\hat \varphi'_\sigma(\tilde k)=\sum U_{\sigma,\sigma'}(\tilde k) \hat \varphi_{\sigma'}(\tilde k)$.  Such states have been extensively studied in the context of velocity selective coherent population trapping, and for each $\tilde k$ the eigenvectors of $H(\tilde k_x)$  are said to form a family of states~\cite{Papoff1992}.  In terms of the associated real-space operators $\hat \psi'_\sigma({\bf r})$ these diagonalized terms of the initial Hamiltonian are
\begin{align}
\hat H_x + \hat H' &=\int d^2 {\bf r} \sum_{\sigma=\pm} \hat\psi'^\dagger_\sigma({\bf r}) E_\sigma\left(-i\frac{\partial}{\partial x},y\right)\hat\psi'_\sigma({\bf r}).\label{TwoLevel1D}
\end{align}
In analogy with the terms ``band'' and ``crystal-momentum'' for particles in a lattice potential, the set of states giving rise to each dispersion curve will be called a ``quasi-band'', and the quantum number $\tilde k_x$ the ``quasi-momentum''.  Here $-i\partial_x$ is the real-space representation of the quasi-momentum $\tilde k_x$.  The symbol $E_\pm\left(-i\partial_x,y\right)$ is a differential operator describing the dispersion of the dressed eigenstates, just as the operator $E_x\left(-i\partial_x,y\right)=(-i\hbar\partial_x-eBy)^2/2m$ describes quadratic dispersion along $\hat x$ of a charged particle in the Landau gauge.

To lowest order in $1/\Omega$ and second order in $\tilde k_x$, $E_\pm(\tilde k_x,y)$ can be evocatively expanded:
\begin{align}
E_\pm& \approx \left(\frac{\Omega}{\Omega\pm4}\right)^{-1}\left(\tilde k_x - \frac{\delta}{4\pm\Omega}\right)^2 + \frac{2\pm\Omega}{2}+\frac{\delta^2(4\pm\Omega)}{4(4+\Omega)^2}.\label{Harmonic}
\end{align}
Atoms in the dressed potential are significantly changed in three ways: (1) the energies of the dressed state atoms are shifted by $\epsilon_\pm = 1\pm \Omega/2$ (the additional energy offset of order $\delta^2$ is relevant to trapping); (2) atoms acquire an effective mass $m^*/m = \Omega / (\Omega\pm4)$, and crucially (3) the center of the dispersion relation is shifted to $eA_x/\hbar\kr = \delta / (\Omega\pm 4)$.  $A_x$ can depend on $y$ by virtue either of a spatial dependence on $\Omega$ (as in Refs.~\cite{Juzeliunas2006,Zhu2006}, which focused on the large $\Omega$ limit), or via $\delta(y) = g\mu_B \Delta B(y)$ as described below.  In either case, the effective Hamiltonian is that of a charged particle in a magnetic field expressed in the Landau gauge.

\begin{figure}[tbp]
\begin{center}
\includegraphics[width=3.375in]{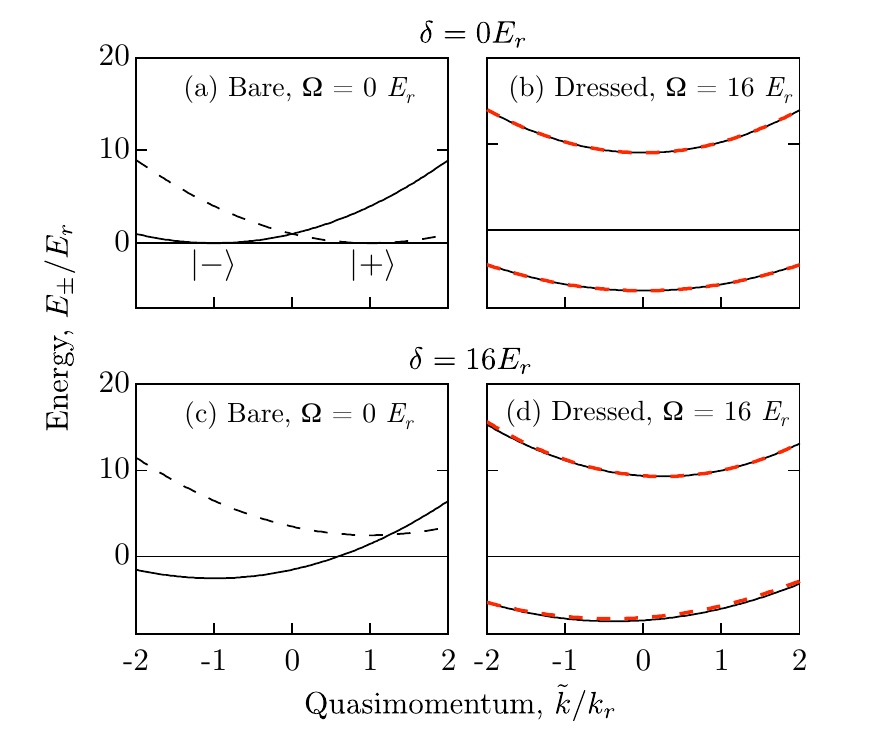}
\end{center}
\caption[Dressed state dispersion for the 2 level system]{
(Color online) Each panel pictures the dressed state dispersion relations for a two level dressed state atoms.  The horizontal axis is quasi-momentum $\tilde k_x$, and the vertical axis is the dressed state energy $E_\pm(\tilde k_x)$.  The black lines are the exact eigenvalues of Eq.~\ref{TwoByTwo}, and the red dashed lines are the analytic approximation.  (a) bare potentials  (undressed) with Raman beams on resonance; (b) dressed potentials ($\Omega=16\Er$, $\delta = 0\Er$); (c) bare potentials (undressed) with Raman beams off resonance ($\delta = 5\Er$); and (d) dressed potentials with Raman beams off resonance ($\Omega=16\Er$, $\delta = 5\Er$).
}
\label{2dressedstates}
\end{figure}

Figure \ref{2dressedstates} shows the dressed state dispersion relations from this model.  Panels a and c show the undressed case ($\Omega=0$) for detuning $\delta=0$ and $5\Er$ respectively.  Panels b and d depict the same detunings, for $\Omega=16\Er$, where the exact results (solid line) are displayed along with the approximate dispersion (red dashed line).  Fig.~\ref{2dressedstates}b then shows the strongly dressed states for large $\Omega$, each of which is symmetric about $\tilde k=0$, when detuned as in panel c, the dispersion is displaced from $\tilde k=0$; when spatially dependent this displacement leads to a non-trivial gauge potential.  In the limit of very small $\Omega$ the dressed curve $E_-(\tilde k_x,y)$ forms a double-well ``potential'' as a function of $k_x$; in a related Raman-coupled system, Bose condensation in such double-well potentials have studied theoretically in Refs.~\cite{Higbie2002,Montina2003,Higbie2004,Stanescu2008}.

It is also possible to treat this problem in the Born-Oppenheimer (BO) approximation in which only $\hat H'$ is diagonalized~\cite{Juzeliunas2006,Zhu2006,Liu2007,Satija2008}.  In this new eigen-basis, $\hat H_x$ has off-diagional terms which are taken to be small, and ignored in the BO approximation.  Such an assumption is valid only when $\Omega^{-1}$ is small, in which case the BO approximation yields a dispersion exactly in the form of Eq.~\ref{Harmonic}, where the crucial terms are $m^*_{\rm BO}/m = 1$ and $e A_{\rm BO} / \hbar k_R = \pm \delta / \sqrt{\delta^2 + \Omega^2}$.  These relations converge to those in Eq.~\ref{Harmonic} for very small $\Omega^{-1}$.  In contrast, the BO approach fails to yield the correct physics in the limit of small coupling, for example never predicting the double-well structure in $E_-(\tilde k_x)$.

\subsection{Effective fields, trapping, and optimization}

\begin{figure}[tbp]
\begin{center}
\includegraphics[width=3.375in]{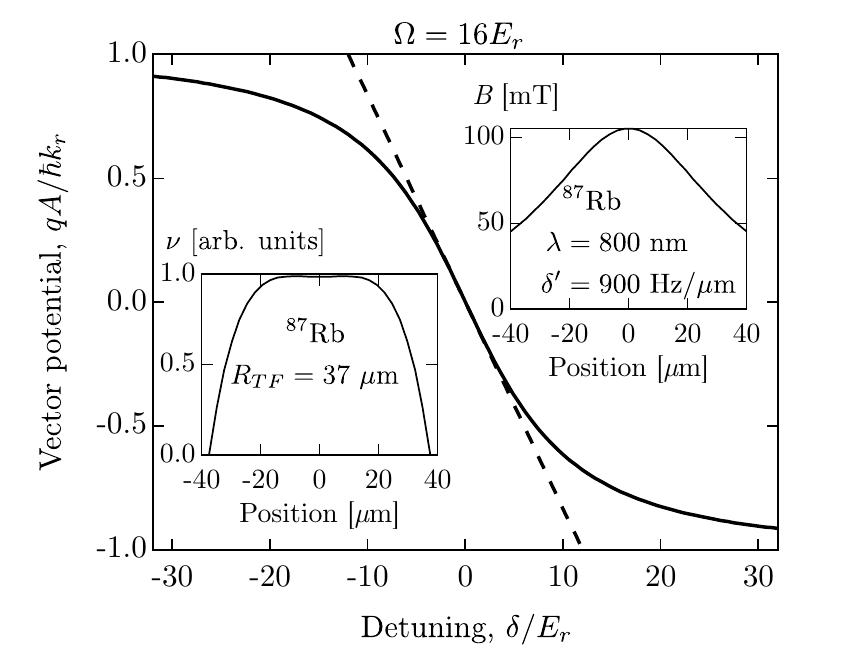}
\end{center}
\caption[Effective vector potential for the 2 level system]{
Effective vector potential $eA_x/\hbar\kr$ versus detuning $\delta$ for $\Omega = 16\Er$.  The solid line is the exact result and the dashed line is the lowest order expansion in $\delta$.  The insets depict predicted quantities relevant to experiment, computed for $\Rb87$ with a detuning gradient $\delta'(y) = 900\Hz/\mu{\rm m}$, and $\lambda = 800\nm$ Raman lasers.  The top inset shows the effective magnetic field $B$, indicating the degree of field inhomogeneity.  The bottom inset shows the filling fraction $\nu$ for a $R_{TF}=37\micron$ BEC chosen so $\nu$ is nearly constant near the systems center.
}
\label{Limitations}
\end{figure}

The analysis of the Raman coupling lead us to a dressed dispersion along $\hat x$, and as is shown below, motion along $\hat y$ is largely unaffected.  When the detuning is made to vary linearly along $\hat y$,  $\delta(y)=\delta' y$, an effective single particle hamiltonian contains a 2D effective vector potential $q{\bf A}/\hbar \kr \approx \left\{\delta' y / (4\pm \Omega),0\right\}$ -- the vector potential for a magnetic field normal to the $\hat x$-$\hat y$ plane expressed in the Landau gauge. The effective magnetic field is $q\beta_z/\hbar\kr \approx \delta' /(4\pm\Omega)$.  Figure~\ref{Limitations} shows the computed vector potential as a function of detuning $\delta$.  As expected, the linear approximation discussed above (dashed line) is only valid for small $\delta$; as a consequence the effective field decreases from its peak value as $\delta$ increases (top inset).

In addition this technique modifies the trapping potential along $\hat y$, i.e, it produces an (unwanted) scalar potential in addition to the vector potential.  When the initial potential $V(x,y)$ is harmonic with trapping frequencies $\omega_x$ and $\omega_y$, the combined potential along $\hat y$ becomes $V_\pm(y) = m (\omega_y^2+(\omega_\pm^*)^2) y^2/2$, where $m (\omega_\pm^*)^2/2 = \delta'^2 (4\pm\Omega)/4(4+\Omega)^2$.

This contribution to the overall trapping potential is not unlike the centripetal term which appears in a rotating frame of reference, where an effective magnetic field $\beta_{\rm rot}$ arrises as well.  In the case of a frame rotating with angular frequency $\Omega_{\rm rot}$, the centripetal term gives rise to a repulsive harmonic term with frequency $\omega_{\rm rot} = (q \beta_{\rm rot} / 2 m)$.  In the present case the scalar trapping frequency can be rewritten in a similar form $\omega_\pm^* = (q \beta / 2 m)\times\left|4\pm\Omega\right|^{3/2}/(4+\Omega)$; the scalar potential may be attractive or repulsive, and it increases in relative importance with increasing $\Omega$.

The effective field generated is inhomogeneous, however, for many physical effects in a magnetic field, such as the Hall effect (quantum and classical) the filling fraction $\nu$, not the magnetic field, is the most relevant parameter.  In the Thomas-Fermi limit the spatial density $n$ of a BEC decreases quadratically from the center of a harmonic trap, at lowest order this can compensate for the decreasing effective field, leading to an extended region of constant $\nu=hn/e\beta$ in the systems center (bottom inset of Fig.~\ref{Limitations}).  This approach is very well suited for incompressible QHE states which will form a shell-structure at constant $\nu$; thus the effective homogeneity is enhanced for a harmonically trapped gas in the presence of a field.  Still, this requires fine-tuning of the system-size for every $\beta$; in the three level case, this fine-tuning restriction is lifted.

\subsection{Additional Coupling}

The preceding calculation omitted motion along $\hat y$, interactions, and an external potential: $\hat H_y$, $\hat H_{\rm int}$, and $\hat V$.  These three complicating terms can be treated easily, starting with the state-independent trapping potential $\hat V$.  The potential can be expressed in terms of dressed quasi-momentum operators via the relations
\begin{align}
\hat V &= \int dy \int\frac{d k_1 d k_2}{(2\pi)^2} \bar V (k_1-k_2,y) \sum_\sigma \hat\phi^\dagger_\sigma(k_1,y) \hat\phi_\sigma(k_2,y) \nonumber\\
 &\approx \int dy \int\frac{d \tilde k_1 d \tilde k_2}{(2\pi)^2} \bar V (\tilde k_1-\tilde k_2,y) \sum_\sigma \hat\varphi'^\dagger_\sigma(\tilde k_1,y) \hat\varphi'_\sigma(\tilde k_2,y)\nonumber\\
 &= \int d^2{\bf r} \sum_\sigma \hat\psi'^\dagger_\sigma({\bf r}) V({\bf r}) \hat\psi'_\sigma({\bf r})\label{DressedPotential}
\end{align}
where $\bar V(\delta k_x,y)$ is the potential Fourier transformed along $\hat x$.  The dressed states experience the same state independent potential as the initial states, however, the small off-diagonal terms of $U(\tilde k_1,y) U^\dagger(\tilde k_2,y)$ together with $\bar V(k_1-k_2,y)$ give transition matrix elements $2 \bar V(k_1-k_2,y) (k_1-k_2)/\Omega$ between dressed states (through second order in $\Omega^{-1}$).  In real space this gives a coupling $\propto \Omega^{-1} \partial_x V(x,y)$, sensible because the coupling term results from deviations from a uniform potential.  Since a typical trap is many tens of wavelengths in extent, and ultracold atoms generally have momenta at or below $k_r$ this coupling term is small, but not in general negligible.

The term describing motion along $\hat y$ also leads to coupling terms between dressed states.  The argument leading to Eq.~\ref{DressedPotential} for $\hat H_y$ gives 
\begin{align}
\hat H_y &\approx -\int d^2{\bf r} \sum_\sigma \hat\psi'^\dagger_\sigma({\bf r}) \partial^2_y \hat \psi'_\sigma({\bf r}),
\end{align}
again having made the approximation $U(\tilde k,y) \partial_y^2 U^\dagger(\tilde k,y)$ $\approx$ $0$; this gives coupling $(2\Omega)^{-1}(\partial_y \delta(y))\partial_y$ at lowest order in $\Omega^{-1}$.  This term results from a breakdown of a BO approximation implicit in the diagonalization of $\hat H_x+\hat H'$ at fixed $y$ leading to Eq.~\ref{TwoLevel1D}~($\hat H'$ depends on $y$ through $\delta(y)$).  This approximation is distinct from the BO approximation alluded to earlier where the coupling Hamiltonian $\hat H'$ alone was diagonalized at fixed $x$ and $y$. 

Finally, the arguments given above also show that the interaction $\hat H_{\rm int}$ leads to a dressed-state independent interaction with the same $g$ with a state-changing coupling term proportional to $g_{\rm 2D} / \Omega$ at order $\Omega^{-1}$.  

Together these allow the construction of the expected ``real space'' Hamiltonian in the basis of localized spin-superposition states $\hat \psi'({\bf r})$.  The BO violating coupling terms discussed above can be treated perturbatively for particles in the lowest quasi-band leading to stable eigenstates, however, transitions for particles starting in higher quasi-bands are energetically allowed and a Fermi's Golden Rule argument thus gives rise to ``decay'' from all but the lowest energy dressed state~\cite{Spielman2006}.

Using the standard argument of a single macroscopically occupied state, the Hamiltonian reduces to the 2D Gross-Pitaevskii equation (GPE)
\begin{align*}
\Big\{\left[E(-i\partial_x,y) - \partial_y^2\right]+\ V({\bf r}) & +\\
(N-1)g_{\rm 2D}\left|\Psi({\bf r})\right|^2\Big\}\Psi({\bf r}) & = \mu\Psi({\bf r}),
\end{align*}
using the $\hat x$ dispersion $E(-i\partial_x,y)$ which {\it parametrically} depends on $y$ from  Eq.~\ref{TwoLevel1D}.  Thus one expects the usual formation of a vortex lattice at small effective fields when the single mode approximation is valid.

\subsection{Limitations}

Naturally, this technique is not without its limitations.  Foremost among them is the range of possible $qA_x/\hbar\kr$ shown in Fig.~\ref{Limitations} where $\Omega = 16\Er$:  while the linear expansion (dashed) is unbounded, the exact vector potential is bounded by $\pm1$.  The reason for this is clear; for example, the hybridized combination of $\ket{+,\tilde k-1}$ and $\ket{-,\tilde k+1}$ cannot give rise to dressed states with minima more positive than $\tilde k = +1$ (where $\ket{1}$ is minimized absent dressing, Fig.~\ref{2dressedstates}c); nor can the minima be more negative than $\tilde k = -1$.

This limitation does not effect the maximum attainable field, only the spatial range over which this field exists.  Specifically, a linear gradient in $\delta(y)$ gives rise to the effective field $\beta_z(y)$ which is subject to $\int_{-\infty}^{\infty}q\beta_z(y)dy=2\hbar\kr$.  This simply states that the vector potential -- bounded by $\pm \hbar\kr/q$ -- is the integral of the magnetic field.  Note however, that along $\hat x$ the region of large $\beta_z$ has no spatial bounds.

A second limitation of this technique is the assumption of strong Raman coupling between the Zeeman split states.  In the alkalis, when the detuning from atomic resonance is large compared to the excited state fine structure the two-photon Raman coupling for $\Delta m_F=\pm 1$ transitions drops as $\Omega\propto\Delta^{-2}$, not $\Delta^{-1}$ as for the AC Stark shift.  As a result, the balance between off-resonant scattering and $\Omega$ is bounded, and cannot be improved by large detuning.  While this is a modest problem for rubidium ($15\nm$ fine structure splitting), it is extremely important for atoms with smaller fine structure splittings: potassium ($\approx 4\nm$) and lithium ($\approx 0.02\nm$).  This issue can be avoided for the two-level case, by using $\Delta m_f = 0$ transitions, e.g., between ground state hyperfine manifolds in the alkalis.

\section{Three-level system}

\begin{figure}[tbp]
\begin{center}
\includegraphics[width=3.375in]{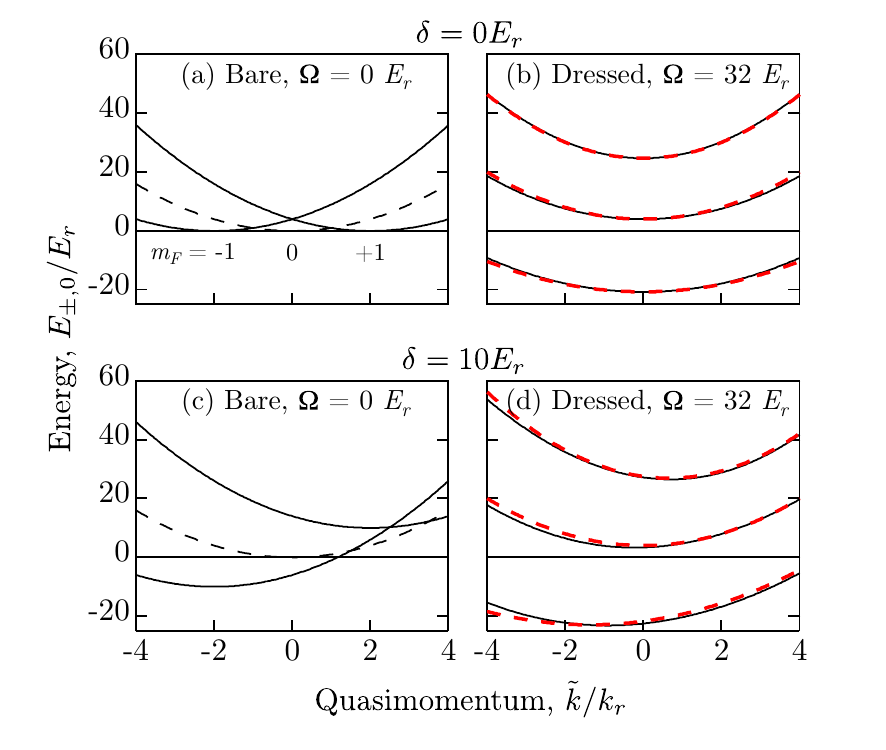}
\end{center}
\caption[Dressed state dispersion for the 3 level system]{
(Color online) Each panel denotes the dressed state dispersion relations $E(\tilde k_x)$ for atoms dressed by counter-propagating Raman beams.  The horizontal axis is quasi-momentum $\tilde k_x$, and the vertical axis is energy in the RWA.  (A) bare potentials  (undressed) with Raman beams on resonance; (B) dressed potentials ($\Omega=32\Er$, $\delta = 0\Er$); (C) bare potentials (undressed) with Raman beams off resonance ($\delta = 5\Er$); and (D) dressed potentials with Raman beams off resonance ($\Omega=32\Er$, $\delta = 10\Er$).
}
\label{dressedstates}
\end{figure}

The range of possible effective vector potentials can be extended by coupling more states, for example the $m_F$ states of a $F>1/2$ manifold in the linear Zeeman regime.  The calculation follows the two-level example above, except for the lack of compact closed-form solutions.  Additional levels extend the range of the vector potential from $\pm \kr$ in the two level case to $\pm 2F\kr$ for arbitrary $F$.

For specificity, consider an optically-trapped system of $\Rb87$ atoms in the $F=1$ mainfold in a small magnetic field which splits the three $m_F$ levels by $g\mu_B |{\bf B}|$ (Fig.~\ref{geometry}c).  The coupling fields can be produced by a pair of far-detuned counter-propagating lasers (aligned normal to the bias field $\bf B$) detuned from each other $\omega_1-\omega_2 = g\mu_B |{ \bf B}| / \hbar - \delta$.  Laser polarizations, $(\sigma_+ + \sigma_-)/\sqrt{2}$ and $\pi$, allow Raman transitions between the hyperfine levels when the detuning $\Delta$ from the excited states is comparable or smaller than the $15\nm$ excited-state fine-structure splitting.

As with the two level case, the 1D Hamiltonian describing motion parallel to the dressing lasers can be made block-diagonal.  The $3\times3$ blocks $H(\tilde k_x)$ describing the three internal states of the $F=1$ manifold are
\begin{equation}
H(\tilde k_x) = \left(
\begin{array}{ccc}
(\tilde k_x - 2)^2 + \delta & \Omega/2 & 0 \\
\Omega/2 & {\tilde k_x}^2 + \epsilon & \Omega/2 \\
0 & \Omega/2 & (\tilde k_x + 2)^2 -\delta
\end{array}\right).\label{ThreeLevelMatrix}
\end{equation}
In this expression, $\delta$ is the detuning of the two photon dressing transition from resonance; $\epsilon$ accounts for any quadratic Zeeman shift; $\Omega$ is the two-photon transition matrix element; and $\tilde k_x$, in units of the recoil momentum $\kr$, is the atomic momentum displaced by a state-dependent term $\tilde k_x = k-2$ for $m_F=-1$, $\tilde k_x = k$ for $m_F=0$, and $\tilde k_x = k+2$ for $m_F=+1$.  When $\Omega\gg8\sqrt{2}$ the three eigenvalues, denoted by $E_{\pm}$ and $E_0$ are approximately:
\begin{align}
E_\pm& \approx \left(\frac{\Omega}{\Omega\pm8\sqrt{2}}\right)^{-1}\left(\tilde k_x - \frac{2\sqrt{2}\delta}{8\sqrt{2}\pm\Omega}\right)^2 + \frac{2\sqrt{2}\pm\Omega}{\sqrt{2}} \label{Expansion3Level} \\ 
E_0 & \approx {\tilde k}^2 + 4. \nonumber
\end{align}
As with the two-level case, the states associated with eigenvalues $E_\pm$ experience an effective vector potential which can be made position-dependent with a spatially varying detuning $\delta$~\footnote{$E_0$ also experiences an effective field, but at higher order in $\Omega^{-1}$.}.  Again, a magnetic field gradient along $\hat{y}$ gives $\delta\propto y$, and generates a uniform effective magnetic field normal to plane spanned by the dressing lasers and real magnetic field $\bf B$. 
\begin{figure}[tbp]
\begin{center}
\includegraphics[width=3.375in]{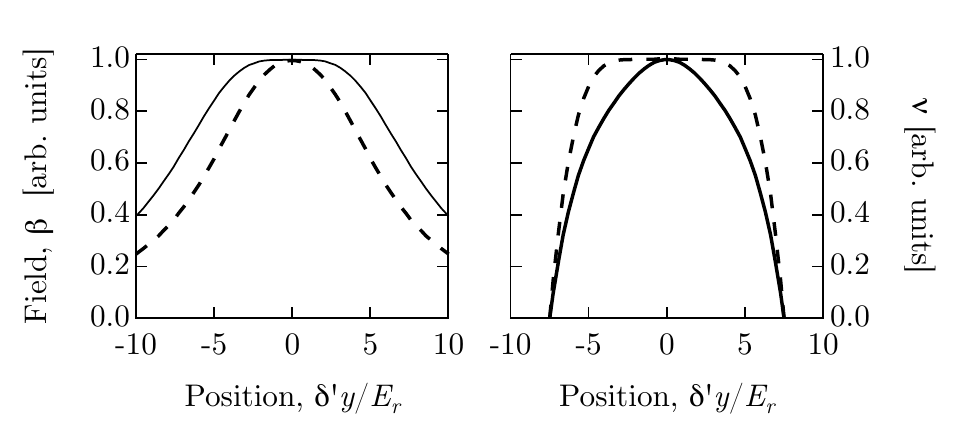}
\end{center}
\caption[Optimized effective fields]{
The left panel shows the optimized effective fields $\beta$ and the right panel the optimized filling fraction $\nu$.  In each plot, $\Omega=16\Er$ and the solid curve shows the case optimized for uniform field ($\epsilon = -1.657\Er$) and the dashed curve for uniform $\nu$ ($\epsilon = -0.064$).  }
\label{Optimizations}
\end{figure}

The resulting magnetic field is inhomogeneous and departs quadratically from its peak value.  For experiments requiring constant filling fraction, $\nu$ can be made approximately uniform by proper selection of the system's Thomas-Fermi radius $R_{TF}$.  Still, some experiments do require a homogenous effective field.  In the lowest energy quasi-band, the proper choice of $\epsilon^* = 4 - \sqrt{2}\Omega/4$ (exact) suppresses the drop-off of the effective magnetic field, leaving terms of order $\delta^4$ and higher ($\epsilon$ results from quadratic Zeeman shifts and is controlled by the bias magnetic field $B$).  For large enough $\Omega$, $\epsilon^* < 0$ corresponding to the physical sign of $\epsilon$ in the $\Rb87$ $F=1$ manifold.  For experiments benefiting from constant $\nu$, a similar analysis shows the filling fraction can be made constant to $O(\delta^4)$ when when $\epsilon^* = 3.2-0.204\Omega$ and when the usual Thomas-Fermi density profile goes to zero at $\delta = 1.00\Omega-8.5$ (Fig.~\ref{Optimizations} shows an example of this optimization as well).

\subsection{Gross-Pitaevskii Equation}

\begin{figure}[tbp]
\begin{center}
\includegraphics[width=3.375in]{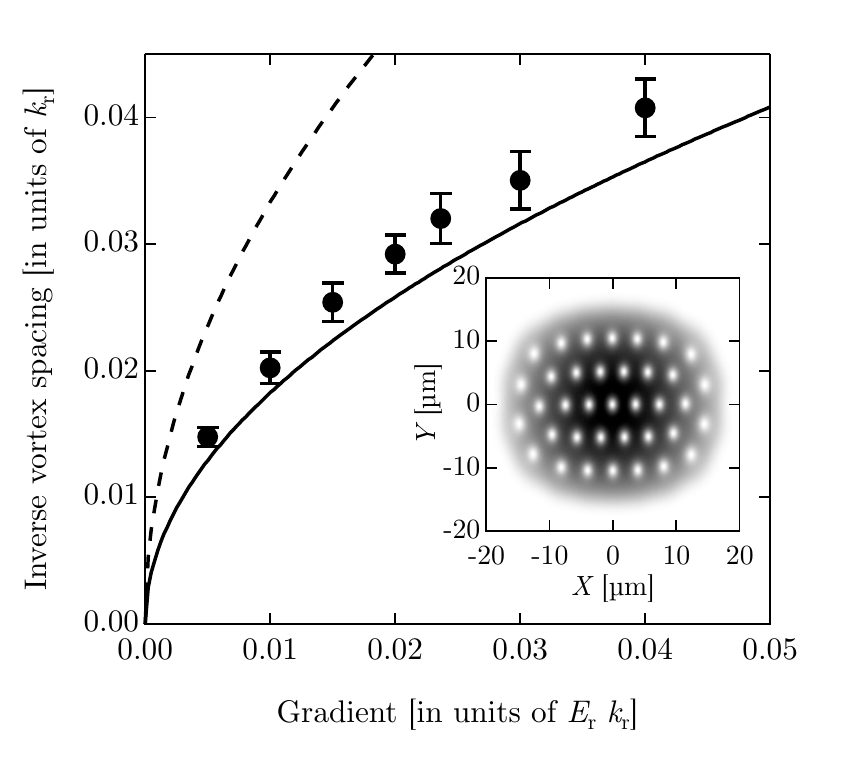}
\end{center}
\caption[Vortex lattice with effective field]{
Inverse vortex spacing versus detuning gradient $\delta'(y)$ in a three level system.  The displayed symbols are obtained by solving the GPE for 3D BEC with $3\times10^5$ to $5\times10^5$ in the presence of an effective magnetic field with a detuning $\delta=0$ at the system's center, and a coupling $\Omega=16\Er$.  The simulation assumes a $10\micron$ Thomas-Fermi radius along $\hat z$, and solves a 2D GPE along the remaining two directions.  The typical vortex spacing is obtained from the Fourier transform of the density distribution $\left|\Psi({\bf r})\right|^2$.  The uncertainties, reflecting the vortex-spacing distribution, are obtained from the half-width of the first peak in the same Fourier transform.  The dashed line is the approximation, Eq.~\ref{Expansion3Level}, and the solid line results from numerical diagonalization of Eq.~\ref{ThreeLevelMatrix}.  Inset: Calculated in-situ density distribution for a detuning gradient $\delta'(y)=0.023\times\Er\kr$ showing the expected vortex lattice structure in a non-symmetric and non-rotating system.
}
\label{vortex}
\end{figure}

The arguments leading to the GPE equation in the two level case remain valid here, and the coupling terms remain of the same order.  A numerical solution to the GPE in the low-field regime is shown in Fig.~\ref{vortex}.  This calculation was performed for the lowest-energy of the three dressed states using the exact dispersion resulting from the numerical diagonalization of Eq.~\ref{ThreeLevelMatrix}.  The computation uses $\Rb87$ parameters and  the experimentally realistic $\Omega=16\Er$.  The inset to Fig.~\ref{vortex} depicts a case with trapping frequencies $\omega_x/2\pi = 10\Hz$ and $\omega_y/2\pi=40\Hz$ (the asymmetry of these terms is partially counteracted by the effective anti-trapping term along $\hat y$ resulting from the zero-offset of the dress state dispersion).  The computed vortex lattice explicitly demonstrates that the approach described above creates an effective field for neutral atoms in a non-rotating frame, even given realistic parameters.  The main panel plots the inverse vortex spacing as a function of gradient directly obtained from the 2D GPE solution (symbols); overlapping these points is a solid line depicting the expected vortex spacing at the systems center (peak effective field) obtained by direct diagonalization of Eq.~\ref{ThreeLevelMatrix}.  The dashed line is the approximate expression from Eq.~\ref{Expansion3Level}.  The formation of the vortex lattice with the correct spacing clearly indicates that this technique does give rise to the expected effective magnetic field.

A counterintuitive reminder of this simulation is that spatially stationary solutions to the dressed-state many body problem exist even when the lowest quasi-band wave-functions intrinsically involve large momentum components and spin-mixtures.  To understand this situation we can consider a more pedestrian example: atoms in an optical lattice.  In this case the systems single particle eigenstates -- Bloch states -- involve only one spin component but are composed of many momentum components each separated by $2\kr$.  In the lowest band of a sinusoidal lattice the $q=0$ Bloch state has no center of mass motion, and instead its many momentum components combine to produce the spatially periodic density modulation characteristic of Bloch states.  In the present case of spin-momentum dressed states, the differing momentum components neither result in center of mass motion, nor in density modulations as with an optical lattice.  Instead, the momentum is associated with a spatially modulated spin texture aligned along $\hat x$.  In both cases, states away from local minima, with non-zero group velocity, do have non-zero mechanical momentum.  The current case differ from the lattice analogy in one substantial way: here the analysis was performed in a frame rotating at the frequency difference between the Raman beams $\Delta_R/h$.  In the rotating frame the spin texture is static, however, in terms of the bare-states the time-dependent phase factors $\exp(i m_F \Delta_R t/\hbar)$ imply that the local orientation of spin texture is rapidly varying.

\section{Conclusions}

Neutral atoms in the presence of suitable coupling laser fields experience effective magnetic fields and the explicitly calculated coupling terms between dressed states are negligible only for atoms in the lowest energy dressed state.  The same effective field effect exists for systems with three or more levels.  This both extends the applicability of the technique, but in addition the additional level increase the spatial extent over which high effective fields can be realized.

I discussed two optimizations: (1) where either the two- or three- level system can be fine-tuned to make the filling fraction constant through third order in displacement along $\hat y$; and (2) where the quadratic Zeeman term in the three-level case allowed the field to be made uniform through third order in $y$ (the same type of reasoning could make $\nu$ constant through order-5 in $y$)~\footnote{These arguments do not include the modification of the density profile resulting from the additional trapping or anti-trapping terms.  Such an optimization is also in principle possible, but it requires details such as the strength of the atom-atom interaction, the number of atoms, and so forth.  These considerations are beyond the scope of this general document.}.

Finally, I showed that an explicit solution to the GPE equation in the presence of an effective gauge potential has the expected vortex lattice.  Using this technique it is possible to generate effective magnetic fields sufficient to enter the FQHE regime, $\nu\lesssim1$ where the GPE is invalid.  For example, in the three level case optimized for uniform field (using Rubidium parameters, $\Omega=16\Er$ and $\epsilon=-1.656\Er$ as in Fig.~\ref{Optimizations}) a gradient of $\delta'(y) \approx h\times 16\kHz/\mu{\rm m}\approx 4.5\Er/\mu{\rm m}$, requires a modest laboratory magnetic field gradient of $2.3\ {\rm T}/{\rm m}$.  This yields an effective field $\beta=4.14\ {\rm mT}$, where the ``effective charge'' was taken to be $e$ (in our technique the product $q\beta$ is defined), and a magnetic length $l_B = \sqrt{\hbar/q\beta}=0.5\micron$.  With a reasonable 2D atom density $n\approx1\micron^{-2}$ the filling fraction is $\nu= h n / q\beta = 1$.  Thus this approach allows experiments to reach the strongly correlated regime with realistic experimental parameters.

% 2.3e-2 T/m = 2.3e-2 G/um -> 16 kHz/um., 1Er = 3.5 kHz, -> 4.5 Er/um
%
% Now A = \hbar k_r * k_0 / e (k_r is there because we use as the dim of k_0)
%
% 

I am deeply appreciative of conversations with V.~Galitski, Y.-J.~Lin, W.~D.~Phillips, J.~V.~Porto, J.~Y.~Vaishnav, C.~A.~R.~Sa~de~Melo, and I.~I.~Satija, and acknowledge the financial support of ONR, DARPAÕs OLE program, and the NSF through the JQI Physics Frontier Center.

\bibliography{main}

\begin{thebibliography}{25}
\expandafter\ifx\csname natexlab\endcsname\relax\def\natexlab#1{#1}\fi
\expandafter\ifx\csname bibnamefont\endcsname\relax
  \def\bibnamefont#1{#1}\fi
\expandafter\ifx\csname bibfnamefont\endcsname\relax
  \def\bibfnamefont#1{#1}\fi
\expandafter\ifx\csname citenamefont\endcsname\relax
  \def\citenamefont#1{#1}\fi
\expandafter\ifx\csname url\endcsname\relax
  \def\url#1{\texttt{#1}}\fi
\expandafter\ifx\csname urlprefix\endcsname\relax\def\urlprefix{URL }\fi
\providecommand{\bibinfo}[2]{#2}
\providecommand{\eprint}[2][]{\url{#2}}

\bibitem[{\citenamefont{Anderson et~al.}(1995)\citenamefont{Anderson, Ensher,
  Matthews, Wieman, and Cornell}}]{Anderson1995}
\bibinfo{author}{\bibfnamefont{M.~H.} \bibnamefont{Anderson}},
  \bibinfo{author}{\bibfnamefont{J.~R.} \bibnamefont{Ensher}},
  \bibinfo{author}{\bibfnamefont{M.~R.} \bibnamefont{Matthews}},
  \bibinfo{author}{\bibfnamefont{C.~E.} \bibnamefont{Wieman}},
  \bibnamefont{and} \bibinfo{author}{\bibfnamefont{E.~A.}
  \bibnamefont{Cornell}}, \bibinfo{journal}{Science}
  \textbf{\bibinfo{volume}{269}} (\bibinfo{year}{1995}).

\bibitem[{\citenamefont{Davis et~al.}(1995)\citenamefont{Davis, Mewes, Andrews,
  van Druten, Durfee, Kurn, and Ketterle}}]{Davis1995a}
\bibinfo{author}{\bibfnamefont{K.~B.} \bibnamefont{Davis}},
  \bibinfo{author}{\bibfnamefont{M.~O.} \bibnamefont{Mewes}},
  \bibinfo{author}{\bibfnamefont{M.~R.} \bibnamefont{Andrews}},
  \bibinfo{author}{\bibfnamefont{N.~J.} \bibnamefont{van Druten}},
  \bibinfo{author}{\bibfnamefont{D.~S.} \bibnamefont{Durfee}},
  \bibinfo{author}{\bibfnamefont{D.~M.} \bibnamefont{Kurn}}, \bibnamefont{and}
  \bibinfo{author}{\bibfnamefont{W.}~\bibnamefont{Ketterle}},
  \bibinfo{journal}{Phys. Rev. Lett.} \textbf{\bibinfo{volume}{75}},
  \bibinfo{pages}{3969} (\bibinfo{year}{1995}).

\bibitem[{\citenamefont{Kinoshita et~al.}(2004)\citenamefont{Kinoshita, Wenger,
  and Weiss}}]{Kinoshita2004}
\bibinfo{author}{\bibfnamefont{T.}~\bibnamefont{Kinoshita}},
  \bibinfo{author}{\bibfnamefont{T.~R.} \bibnamefont{Wenger}},
  \bibnamefont{and} \bibinfo{author}{\bibfnamefont{D.~S.} \bibnamefont{Weiss}},
  \bibinfo{journal}{Science} \textbf{\bibinfo{volume}{305}},
  \bibinfo{pages}{1125} (\bibinfo{year}{2004}).

\bibitem[{\citenamefont{Paredes et~al.}(2004)\citenamefont{Paredes, Widera,
  Murg, Mandel, F{\"o}lling, Cirac, Shlyapnikov, H{\"a}nsch, , and
  Bloch}}]{Paredes2004}
\bibinfo{author}{\bibfnamefont{B.}~\bibnamefont{Paredes}},
  \bibinfo{author}{\bibfnamefont{A.}~\bibnamefont{Widera}},
  \bibinfo{author}{\bibfnamefont{V.}~\bibnamefont{Murg}},
  \bibinfo{author}{\bibfnamefont{O.}~\bibnamefont{Mandel}},
  \bibinfo{author}{\bibfnamefont{S.}~\bibnamefont{F{\"o}lling}},
  \bibinfo{author}{\bibfnamefont{I.}~\bibnamefont{Cirac}},
  \bibinfo{author}{\bibfnamefont{G.~V.} \bibnamefont{Shlyapnikov}},
  \bibinfo{author}{\bibfnamefont{T.~W.} \bibnamefont{H{\"a}nsch}}, ,
  \bibnamefont{and} \bibinfo{author}{\bibfnamefont{I.}~\bibnamefont{Bloch}},
  \bibinfo{journal}{Nature} \textbf{\bibinfo{volume}{429}},
  \bibinfo{pages}{277} (\bibinfo{year}{2004}).

\bibitem[{\citenamefont{Greiner et~al.}(2002)\citenamefont{Greiner, Mandel,
  Esslinger, H{\"a}nsch, and Bloch}}]{Greiner2002}
\bibinfo{author}{\bibfnamefont{M.}~\bibnamefont{Greiner}},
  \bibinfo{author}{\bibfnamefont{O.}~\bibnamefont{Mandel}},
  \bibinfo{author}{\bibfnamefont{T.}~\bibnamefont{Esslinger}},
  \bibinfo{author}{\bibfnamefont{T.}~\bibnamefont{H{\"a}nsch}},
  \bibnamefont{and} \bibinfo{author}{\bibfnamefont{I.}~\bibnamefont{Bloch}},
  \bibinfo{journal}{Nature} \textbf{\bibinfo{volume}{415}}, \bibinfo{pages}{39}
  (\bibinfo{year}{2002}).

\bibitem[{\citenamefont{Hadzibabic et~al.}(2006)\citenamefont{Hadzibabic,
  Kr{\"u}ger, Cheneau, Battelier, and Dalibard}}]{Hadzibabic2006}
\bibinfo{author}{\bibfnamefont{Z.}~\bibnamefont{Hadzibabic}},
  \bibinfo{author}{\bibfnamefont{P.}~\bibnamefont{Kr{\"u}ger}},
  \bibinfo{author}{\bibfnamefont{M.}~\bibnamefont{Cheneau}},
  \bibinfo{author}{\bibfnamefont{B.}~\bibnamefont{Battelier}},
  \bibnamefont{and} \bibinfo{author}{\bibfnamefont{J.}~\bibnamefont{Dalibard}},
  \bibinfo{journal}{Nature} \textbf{\bibinfo{volume}{441}}
  (\bibinfo{year}{2006}).

\bibitem[{\citenamefont{Greiner et~al.}(2005)\citenamefont{Greiner, Regal, and
  Jin}}]{Greiner2005a}
\bibinfo{author}{\bibfnamefont{M.}~\bibnamefont{Greiner}},
  \bibinfo{author}{\bibfnamefont{C.~A.} \bibnamefont{Regal}}, \bibnamefont{and}
  \bibinfo{author}{\bibfnamefont{D.~S.} \bibnamefont{Jin}},
  \bibinfo{journal}{Phys. Rev. Lett.} \textbf{\bibinfo{volume}{94}}
  (\bibinfo{year}{2005}).

\bibitem[{\citenamefont{Nayak et~al.}(2008)\citenamefont{Nayak, Simon, Stern,
  Freedman, and Sarma}}]{Nayak2008}
\bibinfo{author}{\bibfnamefont{C.}~\bibnamefont{Nayak}},
  \bibinfo{author}{\bibfnamefont{S.~H.} \bibnamefont{Simon}},
  \bibinfo{author}{\bibfnamefont{A.}~\bibnamefont{Stern}},
  \bibinfo{author}{\bibfnamefont{M.}~\bibnamefont{Freedman}}, \bibnamefont{and}
  \bibinfo{author}{\bibfnamefont{S.~D.} \bibnamefont{Sarma}},
  \bibinfo{journal}{Reviews of Modern Physics} \textbf{\bibinfo{volume}{80}},
  \bibinfo{pages}{1083} (\bibinfo{year}{2008}).

\bibitem[{\citenamefont{de~Picciotto et~al.}(1997)\citenamefont{de~Picciotto,
  Reznikov, Heiblum, Umansky, Bunin, and Mahalu}}]{de-Picciotto1997}
\bibinfo{author}{\bibfnamefont{R.}~\bibnamefont{de~Picciotto}},
  \bibinfo{author}{\bibfnamefont{M.}~\bibnamefont{Reznikov}},
  \bibinfo{author}{\bibfnamefont{M.}~\bibnamefont{Heiblum}},
  \bibinfo{author}{\bibfnamefont{V.}~\bibnamefont{Umansky}},
  \bibinfo{author}{\bibfnamefont{G.}~\bibnamefont{Bunin}}, \bibnamefont{and}
  \bibinfo{author}{\bibfnamefont{D.}~\bibnamefont{Mahalu}},
  \bibinfo{journal}{Nature} \textbf{\bibinfo{volume}{389}},
  \bibinfo{pages}{162} (\bibinfo{year}{1997}).

\bibitem[{\citenamefont{Camino et~al.}(2007)\citenamefont{Camino, Zhou, and
  Goldman}}]{Camino2007}
\bibinfo{author}{\bibfnamefont{F.~E.} \bibnamefont{Camino}},
  \bibinfo{author}{\bibfnamefont{W.}~\bibnamefont{Zhou}}, \bibnamefont{and}
  \bibinfo{author}{\bibfnamefont{V.~J.} \bibnamefont{Goldman}},
  \bibinfo{journal}{Phys. Rev. Lett.} \textbf{\bibinfo{volume}{98}},
  \bibinfo{pages}{076805} (\bibinfo{year}{2007}).

\bibitem[{\citenamefont{Radu et~al.}(2008)\citenamefont{Radu, Miller, Marcus,
  Kastner, Pfeiffer, and West}}]{Radu2008}
\bibinfo{author}{\bibfnamefont{I.~P.} \bibnamefont{Radu}},
  \bibinfo{author}{\bibfnamefont{J.~B.} \bibnamefont{Miller}},
  \bibinfo{author}{\bibfnamefont{C.~M.} \bibnamefont{Marcus}},
  \bibinfo{author}{\bibfnamefont{M.~A.} \bibnamefont{Kastner}},
  \bibinfo{author}{\bibfnamefont{L.~N.} \bibnamefont{Pfeiffer}},
  \bibnamefont{and} \bibinfo{author}{\bibfnamefont{K.~W.} \bibnamefont{West}},
  \bibinfo{journal}{Science} \textbf{\bibinfo{volume}{320}},
  \bibinfo{pages}{899} (\bibinfo{year}{2008}).

\bibitem[{\citenamefont{Juzeli{\=u}nas
  et~al.}(2006)\citenamefont{Juzeli{\=u}nas, Ruseckas, {\"O}hberg, and
  Fleischhauer}}]{Juzeliunas2006}
\bibinfo{author}{\bibfnamefont{G.}~\bibnamefont{Juzeli{\=u}nas}},
  \bibinfo{author}{\bibfnamefont{J.}~\bibnamefont{Ruseckas}},
  \bibinfo{author}{\bibfnamefont{P.}~\bibnamefont{{\"O}hberg}},
  \bibnamefont{and}
  \bibinfo{author}{\bibfnamefont{M.}~\bibnamefont{Fleischhauer}},
  \bibinfo{journal}{Phys. Rev. A} \textbf{\bibinfo{volume}{73}},
  \bibinfo{pages}{025602} (\bibinfo{year}{2006}).

\bibitem[{\citenamefont{Zhu et~al.}(2006)\citenamefont{Zhu, Fu, Wu, Zhang, and
  Duan}}]{Zhu2006}
\bibinfo{author}{\bibfnamefont{S.-L.} \bibnamefont{Zhu}},
  \bibinfo{author}{\bibfnamefont{H.}~\bibnamefont{Fu}},
  \bibinfo{author}{\bibfnamefont{C.-J.} \bibnamefont{Wu}},
  \bibinfo{author}{\bibfnamefont{S.-C.} \bibnamefont{Zhang}}, \bibnamefont{and}
  \bibinfo{author}{\bibfnamefont{L.-M.} \bibnamefont{Duan}},
  \bibinfo{journal}{Phys. Rev. Lett.} \textbf{\bibinfo{volume}{97}},
  \bibinfo{pages}{240401} (\bibinfo{year}{2006}).

\bibitem[{\citenamefont{Liu et~al.}(2007)\citenamefont{Liu, Liu, Kwek, and
  Oh}}]{Liu2007}
\bibinfo{author}{\bibfnamefont{X.-J.} \bibnamefont{Liu}},
  \bibinfo{author}{\bibfnamefont{X.}~\bibnamefont{Liu}},
  \bibinfo{author}{\bibfnamefont{L.~C.} \bibnamefont{Kwek}}, \bibnamefont{and}
  \bibinfo{author}{\bibfnamefont{C.~H.} \bibnamefont{Oh}},
  \bibinfo{journal}{Phys. Rev. Lett.} \textbf{\bibinfo{volume}{98}},
  \bibinfo{pages}{026602} (\bibinfo{year}{2007}).

\bibitem[{\citenamefont{Cheneau et~al.}(2008)\citenamefont{Cheneau, Rath,
  Yefsah, Gunter, Juzeliunas, and Dalibard}}]{Cheneau2008}
\bibinfo{author}{\bibfnamefont{M.}~\bibnamefont{Cheneau}},
  \bibinfo{author}{\bibfnamefont{S.~P.} \bibnamefont{Rath}},
  \bibinfo{author}{\bibfnamefont{T.}~\bibnamefont{Yefsah}},
  \bibinfo{author}{\bibfnamefont{K.~J.} \bibnamefont{Gunter}},
  \bibinfo{author}{\bibfnamefont{G.}~\bibnamefont{Juzeliunas}},
  \bibnamefont{and} \bibinfo{author}{\bibfnamefont{J.}~\bibnamefont{Dalibard}},
  \bibinfo{journal}{Europhysics Lettes} \textbf{\bibinfo{volume}{83}},
  \bibinfo{pages}{60001 (6pp)} (\bibinfo{year}{2008}).

\bibitem[{\citenamefont{G\"{u}nter et~al.}(2009)\citenamefont{G\"{u}nter,
  Cheneau, Yefsah, Rath, and Dalibard}}]{Gunter2009}
\bibinfo{author}{\bibfnamefont{K.~J.} \bibnamefont{G\"{u}nter}},
  \bibinfo{author}{\bibfnamefont{M.}~\bibnamefont{Cheneau}},
  \bibinfo{author}{\bibfnamefont{T.}~\bibnamefont{Yefsah}},
  \bibinfo{author}{\bibfnamefont{S.~P.} \bibnamefont{Rath}}, \bibnamefont{and}
  \bibinfo{author}{\bibfnamefont{J.}~\bibnamefont{Dalibard}},
  \bibinfo{journal}{Physical Review A (Atomic, Molecular, and Optical Physics)}
  \textbf{\bibinfo{volume}{79}}, \bibinfo{pages}{011604}
  (\bibinfo{year}{2009}).

\bibitem[{\citenamefont{Higbie and Stamper-Kurn}(2002)}]{Higbie2002}
\bibinfo{author}{\bibfnamefont{J.}~\bibnamefont{Higbie}} \bibnamefont{and}
  \bibinfo{author}{\bibfnamefont{D.~M.} \bibnamefont{Stamper-Kurn}},
  \bibinfo{journal}{Phys. Rev. Lett.} \textbf{\bibinfo{volume}{88}},
  \bibinfo{pages}{090401} (\bibinfo{year}{2002}).

\bibitem[{\citenamefont{Lin et~al.}(2009)\citenamefont{Lin, Compton, Perry,
  Phillips, Porto, and Spielman}}]{Lin2009a}
\bibinfo{author}{\bibfnamefont{Y.-J.} \bibnamefont{Lin}},
  \bibinfo{author}{\bibfnamefont{R.~L.} \bibnamefont{Compton}},
  \bibinfo{author}{\bibfnamefont{A.~R.} \bibnamefont{Perry}},
  \bibinfo{author}{\bibfnamefont{W.~D.} \bibnamefont{Phillips}},
  \bibinfo{author}{\bibfnamefont{J.~V.} \bibnamefont{Porto}}, \bibnamefont{and}
  \bibinfo{author}{\bibfnamefont{I.~B.} \bibnamefont{Spielman}},
  \bibinfo{journal}{Physical Review Letters} \textbf{\bibinfo{volume}{102}},
  \bibinfo{pages}{130401} (\bibinfo{year}{2009}).

\bibitem[{\citenamefont{Petrov and Shlyapnikov}(2001)}]{Petrov2001}
\bibinfo{author}{\bibfnamefont{D.~S.} \bibnamefont{Petrov}} \bibnamefont{and}
  \bibinfo{author}{\bibfnamefont{G.~V.} \bibnamefont{Shlyapnikov}},
  \bibinfo{journal}{Phys. Rev. A} \textbf{\bibinfo{volume}{64}},
  \bibinfo{pages}{012706} (\bibinfo{year}{2001}).

\bibitem[{\citenamefont{Papoff et~al.}(1992)\citenamefont{Papoff, Mauri, and
  Arimondo}}]{Papoff1992}
\bibinfo{author}{\bibfnamefont{F.}~\bibnamefont{Papoff}},
  \bibinfo{author}{\bibfnamefont{F.}~\bibnamefont{Mauri}}, \bibnamefont{and}
  \bibinfo{author}{\bibfnamefont{E.}~\bibnamefont{Arimondo}},
  \bibinfo{journal}{Journal of the Optical Society of America B}
  \textbf{\bibinfo{volume}{9}}, \bibinfo{pages}{321} (\bibinfo{year}{1992}).

\bibitem[{\citenamefont{Montina and Arecchi}(2003)}]{Montina2003}
\bibinfo{author}{\bibfnamefont{A.}~\bibnamefont{Montina}} \bibnamefont{and}
  \bibinfo{author}{\bibfnamefont{F.~T.} \bibnamefont{Arecchi}},
  \bibinfo{journal}{Phys. Rev. A} \textbf{\bibinfo{volume}{67}},
  \bibinfo{pages}{023616} (\bibinfo{year}{2003}).

\bibitem[{\citenamefont{Higbie and Stamper-Kurn}(2004)}]{Higbie2004}
\bibinfo{author}{\bibfnamefont{J.}~\bibnamefont{Higbie}} \bibnamefont{and}
  \bibinfo{author}{\bibfnamefont{D.~M.} \bibnamefont{Stamper-Kurn}},
  \bibinfo{journal}{Phys. Rev. A} \textbf{\bibinfo{volume}{69}},
  \bibinfo{pages}{053605} (\bibinfo{year}{2004}).

\bibitem[{\citenamefont{Stanescu et~al.}(2008)\citenamefont{Stanescu, Anderson,
  and Galitski}}]{Stanescu2008}
\bibinfo{author}{\bibfnamefont{T.~D.} \bibnamefont{Stanescu}},
  \bibinfo{author}{\bibfnamefont{B.}~\bibnamefont{Anderson}}, \bibnamefont{and}
  \bibinfo{author}{\bibfnamefont{V.}~\bibnamefont{Galitski}},
  \bibinfo{journal}{Phys. Rev. A} \textbf{\bibinfo{volume}{78}},
  \bibinfo{pages}{023616} (\bibinfo{year}{2008}).

\bibitem[{\citenamefont{Satija et~al.}(2008)\citenamefont{Satija, Dakin,
  Vaishnav, and Clark}}]{Satija2008}
\bibinfo{author}{\bibfnamefont{I.~I.} \bibnamefont{Satija}},
  \bibinfo{author}{\bibfnamefont{D.~C.} \bibnamefont{Dakin}},
  \bibinfo{author}{\bibfnamefont{J.~Y.} \bibnamefont{Vaishnav}},
  \bibnamefont{and} \bibinfo{author}{\bibfnamefont{C.~W.} \bibnamefont{Clark}},
  \bibinfo{journal}{Phys. Rev. A} \textbf{\bibinfo{volume}{77}},
  \bibinfo{pages}{043410} (\bibinfo{year}{2008}).

\bibitem[{\citenamefont{Spielman et~al.}(2006)\citenamefont{Spielman, Johnson,
  Huckans, Fertig, Rolston, Phillips, and Porto}}]{Spielman2006}
\bibinfo{author}{\bibfnamefont{I.~B.} \bibnamefont{Spielman}},
  \bibinfo{author}{\bibfnamefont{P.~R.} \bibnamefont{Johnson}},
  \bibinfo{author}{\bibfnamefont{J.~H.} \bibnamefont{Huckans}},
  \bibinfo{author}{\bibfnamefont{C.~D.} \bibnamefont{Fertig}},
  \bibinfo{author}{\bibfnamefont{S.~L.} \bibnamefont{Rolston}},
  \bibinfo{author}{\bibfnamefont{W.~D.} \bibnamefont{Phillips}},
  \bibnamefont{and} \bibinfo{author}{\bibfnamefont{J.~V.} \bibnamefont{Porto}},
  \bibinfo{journal}{Phys. Rev. A} \textbf{\bibinfo{volume}{73}},
  \bibinfo{pages}{020702(R)} (\bibinfo{year}{2006}).

\end{thebibliography}

\end{document}